\documentclass[sigconf]{acmart}

\usepackage[nolist,printonlyused]{acronym}
\usepackage{listings}
\usepackage{booktabs}
\usepackage{siunitx}
\usepackage{multirow}

\AtBeginDocument{%
  \providecommand\BibTeX{{%
    \normalfont B\kern-0.5em{\scshape i\kern-0.25em b}\kern-0.8em\TeX}}}

\def\see#1{(\cf~\autoref{#1})}

\def\eg{\emph{e.g.}}
\def\ie{\emph{i.e.}}
\def\cf{\emph{cf.}}

\def\rq#1{\textbf{(RQ#1)}}
\def\c#1{\textbf{(C#1)}}

\def\ttplabel{\textit{\aclp*{ttp}}}
\def\infolabel{\textit{Broad information}}
\def\malwarelabel{\textit{Malware used}}
\def\vulnlabel{\textit{Vulnerability targeted}}
\def\relevancelabel{\textit{Relevance}}

\def\ttplabels{\textit{\acsp*{ttp}}}
\def\infolabels{\textit{Info.}}
\def\malwarelabels{\textit{Malw.}}
\def\vulnlabels{\textit{Vulns.}}
\def\relevancelabels{\textit{Rel.}}

\def\threatcrawl{\textsc{ThreatCrawl}}

\renewcommand\footnotetextcopyrightpermission[1]{}

\begin{acronym}
\acro{apt}[APT]{Advanced Persistent Threat}
\acro{bert}[BERT]{Bidirectional Encoder Representations from Transformers}
\acro{bfc}[BFC]{Bootstrapping Focused Crawlers}
\acro{bow}[BoW]{Bag of Words}
\acro{cti}[CTI]{Cyber Threat Intelligence}
\acro{cve}[CVE]{Common Vulnerabilities and Exposures}
\acro{dom}[DOM]{Document Object Model}
\acro{dos}[DOS]{denial-of-service}
\acro{html}[HTML]{HyperText Markup Language}
\acro{ids}[IDS]{Intrusion Detection System}
\acro{ioc}[IOC]{Indicator of Compromise}
\acroplural{ioc}[IOCs]{Indicators of Compromise}
\acro{ml}[ML]{Machine Learning}
\acro{nlp}[NLP]{Natural Language Processing}
\acro{oov}[OOV]{Out-of-Vocabulary}
\acro{osint}[OSINT]{Open Source Intelligence}
\acro{svm}[SVM]{Support Vector Machine}
\acro{ttp}[TTP]{Tactics, Techniques \& Procedures}
\acroplural{ttp}[TTPs]{Tactics, Techniques \& Procedures}
\acro{tf-idf}[TF-IDF]{Term Frequency-Inverse Document Frequency}
\acro{url}[URL]{Uniform Resource Locator}
\acro{pop}[PoP]{Pyramid of Pain}
\acro{sbert}[S-BERT]{Sentence-BERT}
\acro{cvss}[CVSS]{common vulnerability scoring system}
\end{acronym}

\date{\today}

\begin{document}

\title{\threatcrawl: A BERT-based Focused Crawler for the Cybersecurity Domain}

\author{Philipp Kuehn}
\email{kuehn@peasec.tu-darmstadt.de}
\orcid{0000-0002-1739-876X}
\authornote{Corresponding author}

\author{Mike Schmidt}
\email{mike.schmidt@stud.tu-darmstadt.de}

\author{Markus Bayer}
\email{bayer@peasec.tu-darmstadt.de}
\orcid{0000-0002-2040-5609}

\author{Christian Reuter}
\email{reuter@peasec.tu-darmstadt.de}
\orcid{0000-0003-1920-038X}

\affiliation{%
	\institution{Science and Technology for Peace and Security (PEASEC), Technical University of Darmstadt}
	\city{Darmstadt}
	\country{Germany}}

\setcopyright{none}

\begin{abstract}
Publicly available information contains valuable information for \ac{cti}.
This can be used to prevent attacks that have already taken place on other systems.
Ideally, only the initial attack succeeds and all subsequent ones are detected and stopped.
But while there are different standards to exchange this information, a lot of it is shared in articles or blog posts in non-standardized ways.
Manually scanning through multiple online portals and news pages to discover new threats and extracting them is a time-consuming task.
To automize parts of this scanning process, multiple papers propose extractors that use \ac{nlp} to extract \acp{ioc} from documents.
However, while this already solves the problem of extracting the information out of documents, the search for these documents is rarely considered.
In this paper, a new focused crawler is proposed called \threatcrawl{}, which uses \ac{bert}-based models to classify documents and adapt its crawling path dynamically.
While \threatcrawl{} has difficulties to classify the specific type of \ac{osint} named in texts, \eg, \ac{ioc} content, it can successfully find relevant documents and modify its path accordingly.
It yields harvest rates of up to 52\%, which are, to the best of our knowledge, better than the current state of the art.
The results and source code will be made publicly available upon acceptance.
\end{abstract}
\acresetall

\pagestyle{plain}

\maketitle
\acresetall

\section{Introduction}
\label{sec:introduction}

In the field of \ac{cti}, which is a sub-field of \ac{osint}, publicly available information is used to learn from current threats and prevent attackers from using similar \acp{ttp} against other infrastructures.
To do this, \acp{ioc}, \eg\ IP addresses, are shared.
Those \acp{ioc} can then be transformed into detection rules for \acp{ids} to detect and therefore eliminate threats before they can deal any damage~\cite{chrissanders_applied_2013}.

Standards to exchange \acp{ioc} and other \ac{cti}-related information like MISP\footnote{\url{https://www.misp-project.org/}}~\citep{preuveneers_privacypreserving_2022} or STIX\footnote{\url{https://oasis-open.github.io/cti-documentation/stix/intro}} can already be used to publish, import, and export those into \ac{cti} databases and appliances.
Unfortunately, not all information gets published using those standards.
\ac{cti} is often shared in unstructured ways like blog posts or threat reports from security companies or experts~\citep{husari_ttpdrill_2017,ramnani_semiautomated_2017} and scanning through online posts and extracting \acp{ioc} manually is time-consuming.
\citet{hinchy_voice_2022} surveyed 468 full-time security analysts about their work.
66\% stated they spend over half their time on tedious manual work, 64\% believe that more than half of their tasks can be automated, and 64\% stated that they are likely to switch jobs in the next year if there are no modern and automated tools to support them.
This only confirms that developing tools to automate tasks should be a number one priority in the field of \ac{cti}.
Otherwise companies can lose valuable experts which reduces their overall level of security.

Different methods in this regard are already being developed and experimented with.
\ac{ioc} extractors~\citep{liao_acing_2016} and threat detection methods~\citep{lesceller_sonar_2017} are are one example, which automatically extract \acp{ioc} out of usual text.
Others are \ac{cvss} predictors based on open data sources~\citep{elbaz_fighting_2020,kuehn_ovana_2021}.
But to make use of these techniques, appropriate sources need to be used.
Unfortunately, the task of finding those is rarely discussed in this research area.
Usually some specific websites or security blogs are getting selected and then crawled to receive posts to extract the \acp{ioc} from, which leads to limited tools that only use a very small percentage of all the available information.
Since the internet is a very dynamic environment, websites often change, stop publishing new content and new websites get added constantly.
New sources of \ac{cti} would have to be added by researching them or through recommendations and the already used sources have to be reevaluated regularly to check, if they still provide new \ac{cti}-relevant information.

\paragraph{Goal}
\label{par:goal}
Maintaining an up-to-date set of \ac{cti}-relevant websites requires a lot of manual and tedious work.
This leads to a lack of knowledge about personally unknown, but currently active, attack campaigns.
Additionally the manual work leads to less effective and motivated employees, since they have to invest huge amounts of time to search for \ac{cti}.
Those two problems ultimately result in less infrastructure security.
To avoid this, a new and precise approach to find \ac{cti}-relevant websites in a fast and efficient way, based on a focused crawling technique, needs to be developed.
Focused crawlers ``[\dots]~selectively seek out pages that are relevant to a pre-defined set of topics''~\citet{chakrabarti_focused_1999}.
Since this tactic avoids irrelevant regions of the web, it saves hardware as well as network resources.
This work focuses on those unstructured ways to distribute and share \ac{cti}-related information, published in the surface web~\citep{denker_darknets_2019}.

Therefore, this work aims to answer the following research question
\textit{``How can \ac{cti} relevant information be automatically crawled from the clear surface web while adating the crawling path iterativly?''~\rq{}}
Based on this main question, we derive the following sub-questions:
\textit{how can websites in the \ac{cti} domain be efficiently crawled and ranked?~\rq{1}},
\textit{how can websites be categorized automatically as ``\ac{cti}-relevant''?~\rq{2}}, and
\textit{how can those two parts be combined to get an ordered list of \ac{cti}-relevant websites that is as precise as possible?~\rq{3}}

\paragraph{Contributions}
This work proposes a focused crawling pipeline called \threatcrawl{} which combines different techniques of crawling, classifying, and ranking content.
It enables the gathering of specific \ac{cti}-relevant information in the surface web.
The presented crawler uses embedding techniques to create document embeddings of websites during the crawling process to decide, whether sources should be followed~\c{1}.
Such documents are classified according to the type of information and are ranked according to their suitability in the field of \ac{cti}~\c{2}.
The result of this work can be used as a foundation to create fully automated solutions to extract up-to-date \acp{cti}-relevant information.
This could then be leveraged, for example, to create attacker signatures and block them without manual work~\c{3}.

\paragraph{Outline}
Initially, an overview over current works of legal issues in crawling, crawling in general, and different approaches to classify websites~\see{sec:related_work}.
The theoretical concept of \threatcrawl{} is shown in \autoref{sec:theoretical_concept}.
\autoref{sec:implementation} discusses different implementation details.
\autoref{sec:evaluation} shows and discusses different metrics to describe the efficiency of the classifier and the crawler as a whole.
\autoref{sec:conclusion} concludes this work with an overview over the results and findings as well as their implications and future work.

\section{Related Work}
\label{sec:related_work}

This section presents research in the fields of document classification and focused crawling, and outlines the research gap.

\subsection{Document classification}
For the classification we aim at using document embeddings.
Different work in this direction has already been presented.
\citet{peters_deep_2018} and \citet{devlin_bert_2018} propose context-aware word embeddings, \ie, embeddings are generated based on their surrounding context.
\citet{peters_deep_2018} handle \ac{oov} cases by using character convolutions, while \citet{devlin_bert_2018} recursively split tokens up into sub-tokens until each token is recognized.
\citet{devlin_bert_2018} use \ac{bert} that can be trained for specific domains and are often used for a variety of \ac{nlp} tasks due to their flexibility.
\citet{reimers_sentencebert_2019} build upon \ac{bert} by proposing a method to generate sentence embeddings, called \ac{sbert}.
The authors fine-tune the model to create more optimized sentence embeddings.
One work leveraging \ac{sbert} to classify web content is proposed by \citet{tawil_bert_2021}.

\subsection{Focused Crawling \& Threat Extraction}
\citet{wang_focused_2010} present a \ac{tf-idf} algorithm in combination with a Naive Bayes classifier to find pages relevant to a specific topic.
\citet{liu_focused_2023} extend this work and use a semantic disambiguation graph to remove ambiguation terms that are irrelevant to the topic.
They then use a semantic vector space model to compute the similarity between texts, using \ac{tf-idf} as weights.
A work searching for data lakes is presented by \citet{zhang_dsdd_2021}.
It aims to fill the gap between datasets, annotated to be found by the Google Dataset Search and openly available datasets missing this metadata.
\citet{koloveas_intime_2021} propose a \ac{ml}-based integrated framework for acquiring, analysing, managing, and sharing \ac{cti}-related information.
For the present work, especially the part of crawling and classifying is of interest.
The authors propose a two-step approach, first, a very broad crawl and later a filtering step for the relevant content.
For relevance classification Gensim~\citep{rehurek_software_2010} is used, an open-source implementation of Word2Vec~\citep{mikolov_efficient_2013}.
When classifying a downloaded document, all words present in the topic vocabulary are extracted, embedded, and summed up.
The resulting vector is then compared to the topic vectors and a cosine similarity between those is computed to receive a relevance score.

Other proposals in \ac{cti} research suggest to focus on Twitter, since it is an easy access to cyber threat information~\citep{tundis_automated_2020,tundis_featuredriven_2022,riebe_cysecalert_2021,shin_twiti_2021}.
Others use a set of pre-known websites~\citep{liao_acing_2016}.
While both ways result in valid studies, they are based on pre-known information, either in forms of twitter-handles or websites, without the possibility to widen the view

\subsection{Research Gap}
Current works that combine web crawling, classifying, and page ranking techniques for \ac{cti} into a single pipeline are rare.
\citet{tawil_bert_2021} describe a crawler that uses embeddings~(\ac{sbert}) to assign labels to downloaded documents.
With this approach crawling and classifying are two fully independent processes, \ie, everything gets crawled and is classified only afterwards.
\citet{koloveas_intime_2021} present another two-step approach for crawling and classifying their documents.
The difference in their approach is, that they use a focused crawler, which filters certain documents, but still performs a very broad search.
This leads to a very low relevance rate~($\approx 1\%$) of downloaded websites~\citep{koloveas_crawler_2019}.
Hence, the actual benefits of focused crawling~(saving hardware and network resources) are diminished.
It should be mentioned that \citet{koloveas_intime_2021} also implement additional crawlers that are specifically tailored for fixed websites.
Thus, an open question is the development of a generalized one-step focused crawling approach, that combines the crawling, classifying, and ranking of \ac{cti}-relevant content.

\section{\threatcrawl{}}
\label{sec:theoretical_concept}
This section presents the architecture for \threatcrawl{}.
First, we gloss the abstract concept of relevance, followed by legal issues when dealing with web crawling.
The architectural design decisions are presented afterwards, followed by the classifications process and the crawler seed.
Details concerning the implementation and datasets are shown in \autoref{sec:implementation}.

\subsection{Relevance}
\label{subsec:what_ist_relevant}
Research is trying to capture a definition of \ac{cti} relevance to empirically support automated \ac{cti} research~\citep{kuehn_notion_2022}.
\ac{cti} can be defined as ``the set of data collected, assessed and applied regarding security threats, threat actors, exploits, malware, vulnerabilities and compromise indicators''~\cite{shackleford_whos_2015}.

\begin{table}
	\centering\small
	\begin{tabular}{@{} cp{.2\linewidth}p{.3\linewidth}p{.3\linewidth} @{}}
		\toprule
		             & \textbf{\citeauthor{lee_2020_2020}} & \textbf{\citeauthor{brown_sans_2021}}                         & \textbf{\citeauthor{brown_sans_2022}}                         \\ \midrule
		\textbf{\#1} & \acsp*{ioc}                         & Information about vulnerabilities being targeted by attackers & Detailed information about malware being used in attacks      \\ \addlinespace
		\textbf{\#2} & \acsp{ttp}                          & Detailed information about malware being used in attacks      & Information about vulnerabilities being targeted by attackers \\ \addlinespace
		\textbf{\#3} & Strategic analysis of the adversary & \acfp*{ioc}                                                   & Broad information about attacker trends                       \\
		\bottomrule
	\end{tabular}
	\caption{Top 3 most useful \acf*{cti} types according to SANS CTI survey~\citep{lee_2020_2020,brown_sans_2021,brown_sans_2022}.}
	\label{tab:most_useful_cti_types}
\end{table}

Likewise, \citet{bianco_pyramid_2014} defines the \ac{pop}~\see{fig:pyramid_of_pain}, where ranks and \acs{ioc}-types correspond with the effectiveness on preventing attacks.
While hash values are very easy to collect and detect, their effectiveness is also very limited.
\acp{ttp} on the other hand are hard to collect and detect, but being able to identify them is very effective to detect a certain attacker.

\begin{figure}
	\centering
	\includegraphics[width=0.7\linewidth]{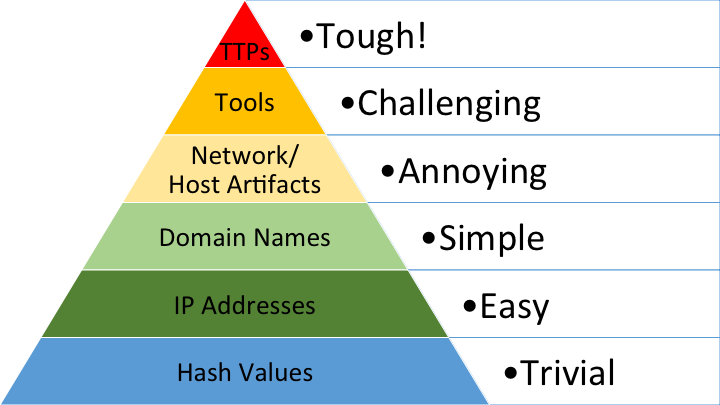}
	\caption{\acl*{pop} by~\citet{bianco_pyramid_2014}}
	\label{fig:pyramid_of_pain}
\end{figure}

The SANS Institute\footnote{\url{https://www.sans.org}} conducts an annual \ac{cti} survey where they interview security professionals from various organizations.
The top 3 most valuable and useful \ac{cti} types are shown in \autoref{tab:most_useful_cti_types}.
In 2020 the most useful type was \acp{ioc}.
In 2021 those were only on 3$^\mathit{rd}$~place and in 2022 they were moved to 5$^\mathit{th}$~place.
Instead of this very specific \ac{cti} type, more general types moved into the most useful types, having "Broad information about attacker trends" in 3$^\mathit{rd}$~place in 2022.
Looking again at the \ac{pop}, the focus is shifting from the lower levels towards the upper levels of the pyramid.

When testing different approaches, \acp{ioc} turned out to be bad indicators.
While just detecting those kind of information can be done very easily with regular expressions, \acp{ioc} like email addresses are often not related to any kind of attacker and hence are not relevant in this context.
Therefore, the top 4 most useful types of \ac{cti} of the latest SANS survey will be used in this work, namely:
\begin{enumerate}
	\item Detailed information about used malware in attacks~(\malwarelabel)
	\item Information about targeted vulnerabilities by attackers~(\vulnlabel)
	\item Broad information about attacker trends~(\infolabel)
	\item \ttplabel
\end{enumerate}

For the context of this work, every website that contains information belonging to one or more of the above types is considered \ac{cti}-relevant.

\subsection{Legal Requirements}
When crawling the internet, legal requirements must be followed.
\citet{krotov_legality_2020} discuss in their paper different aspects of crawling regarding legality and ethics.
The authors criticize that despite a lot of research in that area, legal and ethical issues often get ignored.
In terms of legality, web scraping is still a grey area.
Websites can define \textit{terms of use}, but since users need to agree to them specifically, this does not really apply to automated tools.
Other legal issues include the use and processing of copyrighted material and the purpose of web scraping.
While a lot of work can be categorized as ``fair use'', any fraudulent or prohibited use of the collected data is usually still illegal by law.
In terms of ethics, the privacy of users appearing in the collected data should be respected.
User-generated content like reviews is being collected without considering the implications for those users.
Additionally, crawling should not be used to create products that directly compete with the content that was crawled from other sources.

To fully comply to all terms of use, a crawler first would have to locate, download, and analyze them, which results in a incredibly complex task, considering that each owner can define those differently.
To still enable website administrators to block unwanted crawlers, the Robots Exclusion Protocol was developed\footnote{\url{https://www.rfc-editor.org/rfc/rfc9309.html}}.
Our implementation should respect this protocol and act as defensively as possible.

\subsection{Architecture}
\label{subsec:architecture}

\begin{figure}
	\includegraphics[width=\linewidth]{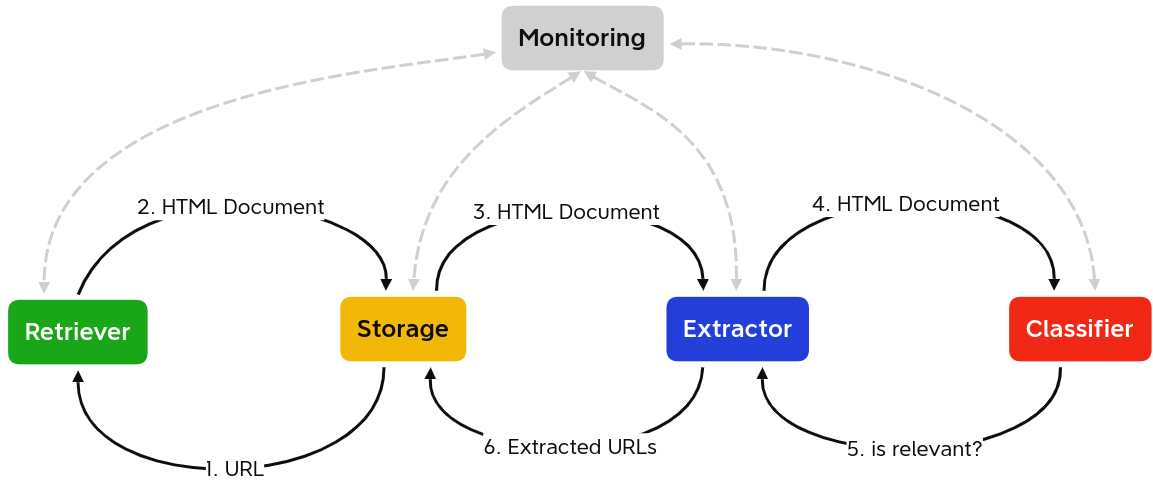}
	\caption{The architecture}
	\label{fig:architecture}
\end{figure}

While there are different crawlers available, like the ACHE Focused Crawler\footnote{\url{https://github.com/VIDA-NYU/ache/}}~\citet{koloveas_intime_2021} or \ac{bert} based Topic-Specific Crawler~\citep{tawil_bert_2021}, none of those offer to use a custom embedding mechanism for documents.
Hence, we opt for a new architecture called \threatcrawl{}, which implements a focused crawler in one step rather the presented two-step approaches~\citep{koloveas_intime_2021,tawil_bert_2021}.
\autoref{fig:architecture} depicts the basic design of the developed solution.
\threatcrawl{} is divided into five logical modules, namely the storage, retriever, extractor, classifier, and monitor.

\paragraph{Storage}
\label{subsec:storage_module}
The storage module contains the \ac{url} queue, crawled \acp{url}, the robots.txt database, domain timers, and databases for the processed and unprocessed \ac{html} files.
The \ac{url} queue contains all \acp{url} that need to be requested.
The retriever processes this queue while the extractor puts new ones into the queue.
All \acp{url} that have been visited will be saved in this list to avoid multiple and unnecessary requests to the same resource.

The robots.txt database manages all tasks related to ``robots.txt'' files that the retriever downloads from different domains, including access rules and crawl-delays.
If no crawl-delay is defined, a default is used to ensure a defensive crawling process.
Additionally, this component enforces the access rules defined by the crawled server in the robots.txt.

Lastly, the storage module saves unprocessed and processed \ac{html} documents that are requested by the retriever module and processed by the extractor module, respectively.
The processed \ac{html} documents are enriched with the extracted \acp{url} and the result of the classification process.

\paragraph{Retriever}
\label{subsec:retriever_module}

The retriever is requests \ac{html} documents and handles all outgoing connections.
It uses the \ac{url} queue and robots.txt database to ensure a proper crawling access and delay.
After the request the received \ac{html} document is saved in the unprocessed \ac{html} database.
To give administrators of crawled webpages more information about \threatcrawl{}, a link is embedded in the user-agent with a description and instructions to prevent it from accessing their site.

\paragraph{Extractor Module}
\label{subsec:extractor_module}

The extractor scans unprocessed \ac{html} documents and extracts all valid \acp{url}\footnote{\url{https://developers.google.com/search/docs/crawling-indexing/robots-meta-tag}}.
It parses those documents and hands it over to the relevance classifier.
If the document is marked as irrelevant,  no further processing is done.
Otherwise the extractor extracts all valid \textit{href} attributes~(\eg, removal of mailto-link, check for \texttt{nofollow} tag) of anchor objects in the \ac{html}'s \ac{dom} and transformes them into absolute \acp{url}.
The remaining addresses are compared to a blacklist~(\eg, youtube links), added to the \ac{url} queue, and the processed document is saved in the \ac{html} database.

\paragraph{Classifier}
\label{subsec:classifier_module}

The classifier is responsible whether \ac{html} are relevant.
Websites contain a lot of irrelevant elements~(\eg, footers, headers).
Hence, the classifier removes all of those elements and only extract the main content of the received documents.
After the extraction, a document embedding is created that represents the content of the document.
This embedding is compared to a ground truth vector~\see{subsec:classification} to identify whether the document is relevant.

\paragraph{Monitoring Module}
\label{subsec:monitoring_module}

The monitor module stores the state of each retriever and extractor thread.
Every time a thread changes its state, it communicates that transition to this module.
Additionally, in case any critical error appears during the execution of \threatcrawl{}, the monitor module can stop all threads immediately.

\subsection{Classification}
\label{subsec:classification}

The process of classifying can be split into two parts.
In the first part, ground truth vectors have to be generated to act as reference vectors for each label.
By splitting up the label ``relevant'' into multiple sub-labels, like the ones mentioned in \autoref{subsec:what_ist_relevant}, we can classify the content even more precisely.
The second part is the classification and labeling of single documents, which is done when running \threatcrawl{}.
An embedding for each requested document is created and compared to the generated ground truth vectors.

\paragraph{Ground Truth Vectors}
\label{subsub:generate_ground_truth_vectors}
Contextualized embeddings are used to create ground truth vectors for each label.
The sum of all token embeddings for a document generates the document embedding~\citep{koloveas_crawler_2019}.
To generate a ground truth vector for a label, every document with that label in the train dataset is embedded and added up to one vector, which is then normalized.

\paragraph{Calculating Allowed Distance}
With the ground truth vector, the angle of every document of that label to the ground truth vector is calculated to approximate the \textit{allowed distance}.
This is done by using the cosine similarity and either use the maximum of all resulting distances or the average over all distances per label.
The lower the angle, the higher the semantic similarity between two embeddings.
The ground truth vector and its allowed distance is saved as output of the training phase for each label.

\begin{figure}
	\centering
	\includegraphics[width=0.9\linewidth]{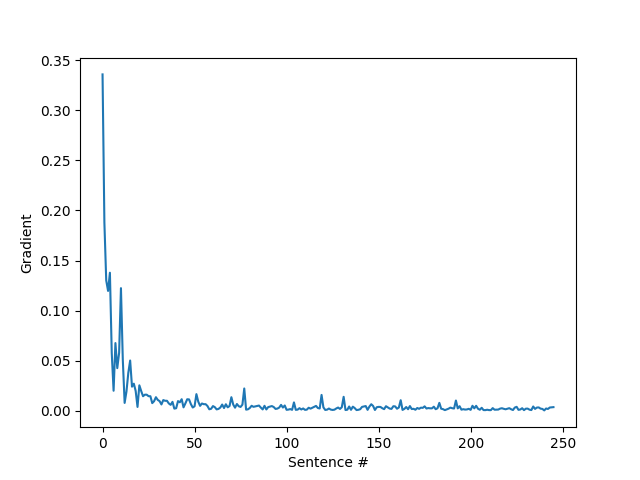}
	\caption{Example of the change of gradients over the course of a document}
	\label{fig:sentence_gradients_single_document}
\end{figure}

\paragraph{Adaptive Sentence Usage}
The calculation time of ground truth vectors varies with the number of sentences per document.
To reduce the processing time when generating embeddings, a predefined amount of sentences could be used, but this number depends on the dataset.
Other metrics like median or average over all documents are not suitable, since the amount of sentences per document varies greatly, which can be seen in the analysis of the dataset~\see{sec:evaluation}.
While generating a the document embedding, the change of the embedding after each embedded sentence is monitored by measuring the distance between before and after each sentence.
An example on how these gradients change is shown in  \autoref{fig:sentence_gradients_single_document}.
It can be seen that the first sentences clearly have the most influence, while all later sentences have significantly less impact on the document embedding.
To boost the performance of the classification process, an adaptive sentence usage can be calculated.
To do that, a fixed gradient value is chosen.
As soon as the gradient after each newly added sentence embedding is below that value, the embedding process can be stopped, since additional sentences will not change the document embedding significantly anymore.

This adds a preparation phase to the generation of the ground truth vectors, where the ideal amount of sentences is calculated for every single document in the training set.
This number is then averaged over all documents, resulting in the adaptive sentence usage that then is used in the ground truth generation.
While this roughly doubles the processing time while training, it reduces the processing time when classifying documents later.

The result of the training process is therefore a ground truth vector, an allowed distance, and an adaptive amount of sentences per label.

\paragraph{Classifying Single Documents}
\label{subsubsec:classify_single_documents}
When running \threatcrawl{}, each received document is labeled by calculating the document embedding as described previously.
After that, the distance to each ground truth vector is calculated.
If the distance is within than the allowed distance, the document can be labeled according to that ground truth vector.

\begin{figure}
	\centering
	\includegraphics[width=0.7\linewidth]{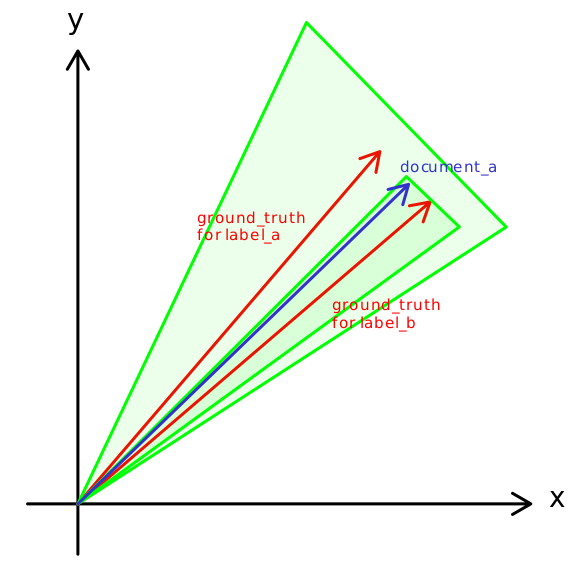}
	\caption{Illustration of the relative distance}
	\label{fig:relative_distance}
\end{figure}

Unfortunately, when having overlapping ground truth vectors, the normal distance will lead to wrong labels.
In \autoref{fig:relative_distance} two different ground truth vectors~(red) and their allowed distances~(green) are shown.
The embedding for \textit{document\_a} is inside the allowed distance for both labels, but closer to \textit{label\_b}.
Relatively to the ground label and and its allowed distance, the document should be labeled with \textit{label\_a}, but by using the normal distance it will be labeled with \textit{label\_b}.
To avoid this, additionally to the distance, the \textit{relative distance} will also be calculated by dividing the distance by the allowed distance.

The relative distance can also be used as an inverted relevance score.
The lower the relative distance, the more relevant a document is.

\begin{figure}
	\centering
	\includegraphics[width=\linewidth]{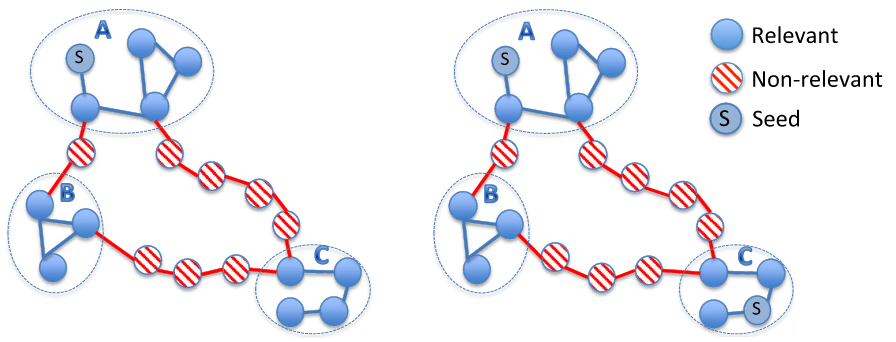}
	\caption{Example of isolated components~\cite{vieira_finding_2016}}
	\label{fig:isolated_components}
\end{figure}

\subsection{Choosing the Right Seed}
\label{subsubsec:choosing_right_seed}
When a crawler is initially started, it needs to be given a initial seed to start processing.
Choosing the right seed for a focused crawler is crucial for its performance~\citep{vieira_finding_2016}.
Since focused crawlers only look for relevant information, their performance can suffer if there are no connections between different clusters of relevant information like depicted in \autoref{fig:isolated_components}~\citep{vieira_finding_2016}.
Clusters B and C will never be discovered by the crawler since there are no connections between A and B/C.
This problem can be solved or mitigated by choosing a more diverse seed.
In the right part of \autoref{fig:isolated_components} we can see, that now a part of C is also used as seed, which enables the crawler to find more clusters of relevant information.
\citet{vieira_finding_2016} find, that a seed has to be constructed automatically, since using resources like web directories will often lead to ineffective seeds for topics that are underrepresented or do not exist in those directories.
They propose a framework called \ac{bfc} that iteratively performs queries to search engines and classifies the returned \acp{url}, to find the ones that are more likely to be suitable seeds.
That way the authors create a high number of appropriate and diverse \acp{url} that improve the performance of a focused crawler.

\section{Implementation}
\label{sec:implementation}
In this section, specific details about the implementation of all concepts mentioned in \autoref{sec:theoretical_concept} are given.

\subsection{Used Libraries}
As mentioned in \autoref{subsec:architecture}, a new crawler called \threatcrawl{} had to be developed.
To make \threatcrawl{} as simple and lightweight as possible, native Python was primarily used.
The following libraries/modules are the only ones that are used additionally to the built-in ones:
\begin{itemize}
	\item Beautiful Soup\footnote{\url{https://www.crummy.com/software/BeautifulSoup/}}: library for parsing traversing and manipulating \ac{html} content and documents
	\item Protego\footnote{\url{https://github.com/scrapy/protego}}: library for parsing "robots.txt" files
	\item NumPy\footnote{\url{https://numpy.org/}}: library for scientific computing
	\item PyTorch\footnote{\url{https://pytorch.org/}}: \ac{ml} framework
	\item Transformers\footnote{\url{https://pypi.org/project/transformers/}}: library for loading and handling tokenizers and \ac{ml} models
	\item Trafilatura\footnote{\url{https://trafilatura.readthedocs.io/en/latest/}}: library for discovering and extracting web content
\end{itemize}

To extract the main content of a webpage \threatcrawl{} uses Trafilatura~\cite{barbaresi_trafilatura_2021}, which provides a simple function that extracts the main content of an \ac{html} document and returns it as parsed and plain text.

A blacklist for domains is essential to block common domains like Youtube or Facebook.
Websites like these are widespread but have no benefit when searching for \ac{cti} relevant information, especially when focusing on text documents.
Even worse, since a lot of them hide their content behind logins, they can not even be considered part of the Clear Web.
Unfortunately, there is no realistic list of all websites that do not belong to the Clear Web.
Therefore, the top 100 most used websites were used as a starting point to filter websites, that usually do not serve any content relevant for this work.

Since no appropriate dataset was available, a new one had to be created.
To do so, multiple blogs and websites containing articles with \ac{cti}-relevant information as defined in \autoref{subsec:what_ist_relevant} were selected.
This resulted in a dataset of 259 \acp{url}.
Out of these 259 \acp{url}, 153 were labeled \emph{relevant} and 106 \emph{not relevant}.
To enable a more precise classification, the label \emph{relevant} was divided into sub-labels.
The final dataset therefore consists of the following documents per label:
\begin{itemize}
	\item Relevant: 153
	      \begin{itemize}
		      \item \ttplabel: 55
		      \item \infolabel: 24
		      \item \malwarelabel: 36
		      \item \vulnlabel: 38
	      \end{itemize}
	\item Not Relevant: 106
\end{itemize}

\begin{figure}
	\centering
	\includegraphics[width=\linewidth]{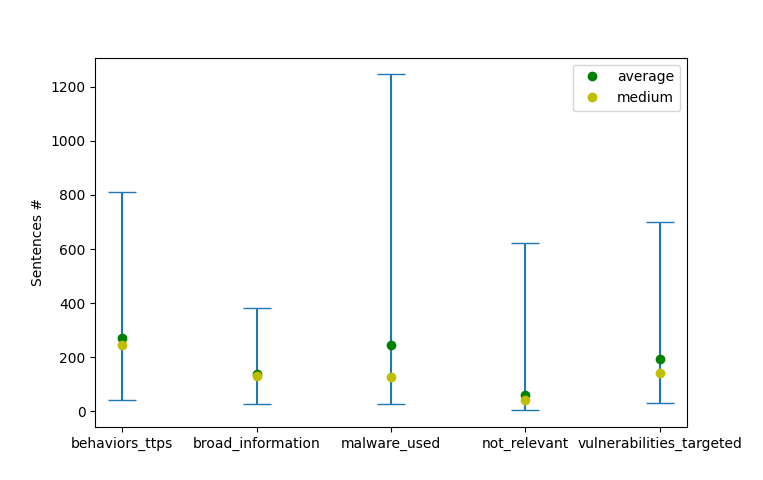}
	\caption{Sentences per label in the dataset}
	\label{fig:dataset}
\end{figure}

It is important to note, that the sub-labels often overlap.
Articles describing \acp{ttp}, for example, also contain information about used malware and the vulnerabilities that they target.
This can also be seen in the analysis of the most important sentences per label~\see{subsec:eval_classifier}.
In \autoref{fig:dataset} the amount of sentences extracted per label can be seen.

\subsection{Contextualized Embeddings}
To create the contextualized embeddings, a pre-trained \ac{bert} model was used, called CySecBERT~\cite{bayer_cysecbert_2024}.
CySecBERT is a fine-tuned version of the \emph{bert-base-uncased} model and was trained on 528M tokens collected from content in the IT security environment.
The resulting model has 12 hidden layers with 768 hidden units in each layer.

Before embedding a sentence, it has to be pre-processed by tokenizing it and adding the start token and the seperator token at the end:
\[\mathrm{Input}(sentence)=[\mathrm{CLS}]\;\mathrm{Tokenize}(sentence)\;\mathrm{[SEP]}\]

Since the input layer can only handle a maximum of 512 tokens, every token over that limit has to be removed.
\citet{devlin_bert_2018} state that concatenating the last four hidden layers achieves high scores even without fine-tuning the model.
Using the result of the last four layers, a contextualized embedding for each token in the sentence is created.
Since each layer contains 768 hidden units, each embedding has a length of 4 * 768 = 3027.
These token vectors can then be added up to create sentence vectors, which afterwards can be added up to create one single document vector.

\paragraph{Adaptive Amount of Sentences}
To calculate the amount of sentences, sentence gradients for all relevant documents in the dataset were calculated~\see{fig:sentence_gradients}.
A value of 0.02 was chosen as the minimum that has to be reached, before no more significant value will be gained by considering more sentences.

\begin{figure}
	\centering
	\includegraphics[width=\linewidth]{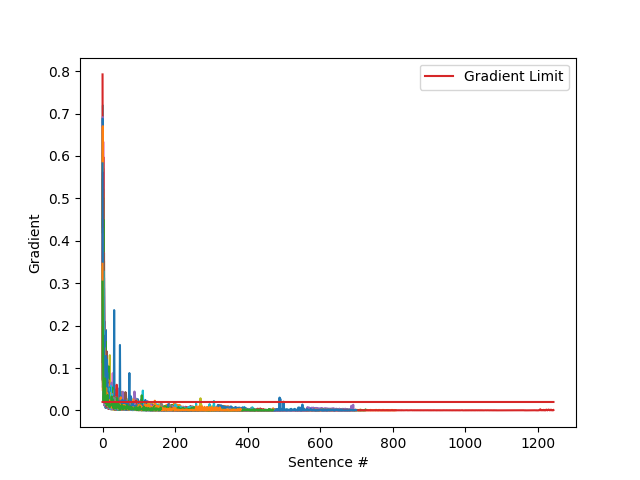}
	\caption{The sentence gradients}
	\label{fig:sentence_gradients}
\end{figure}

\paragraph{Choosing the Right Seed}
To maintain reproducibility, a fixed seed was used.
The seed consists of 51 \acp{url} with a huge variety between websites that typically host \ac{cti} relevant content and websites that do not.
Some of the domains in the seed can also be found in the dataset and some are completely new.

\section{Evaluation}
\label{sec:evaluation}

In this section the evaluation method will be described~\see{subsec:evalmethod}, followed by an evaluation of the classification module~\see{subsec:eval_classifier} in the context of correct classification of documents.
Lastly, \threatcrawl{} will be evaluated in the context of a focused crawler as a whole, \eg, by evaluating its harvest rate~\see{subsec:eval_crawl_bot}.

\subsection{Evaluation Method}
\label{subsec:evalmethod}

By analyzing the most important sentences, characteristics for each label can be observed.
To extract the most important sentences in each document, the embedding of each sentence in a document is compared to the calculated embedding of the whole document.
The sentence with the highest similarity is the one sentence that best captures the essence of the document.

To evaluate the classifier, a K-fold cross-validation, with K set to 5, was used, which is especially effective on sparse datasets~\citep{hastie_elements_2009}.
To evaluate the CySecBERT model~\citep{bayer_cysecbert_2024}, multiple evaluations with 2 different parameters were conducted.
The first parameter is the \textit{amount of sentences per document} when creating the document embedding.
As described in \autoref{subsub:generate_ground_truth_vectors}, instead of a fixed amount, an adaptive amount can be used~(set as default).
The second parameter is the \textit{calculation of allowed distance}, which can either be the average or the maximum distance over all distances for each label~\see{subsub:generate_ground_truth_vectors}~(average set as default).

For \threatcrawl{}'s performance the ratio of downloaded documents to downloaded relevant documents is measured, yielding the focusing efficiency.
Additionally, an analysis of the resulting list of relevant documents and their relationship to each other is done.

All calculations were done on a Intel® Core™ i7-8550U processor with 4 GHz.

\subsection{Evaluation of the Classification Module}
\label{subsec:eval_classifier}

\begin{table*}[]
    \centering
    \small
    \begin{tabular}{@{} r p{.94\linewidth} @{}}
    \toprule
    \textit{Rank} & \textbf{\ttplabel} \\
    \midrule
        1 & Instead, the threat actor leveraged a misconfiguration in GitHub repositories to get code execution and initial access on thousands of hosts across what are likely multiple victim environments worldwide \\
        2 & Based on the overlaps between UNC2165 and Evil Corp, we assess with high confidence that these actors have shifted away from using exclusive ransomware variants to LOCKBIT—a well-known ransomware as a service (RaaS)—in their operations, likely to hinder attribution efforts in order to evade sanctions \\
        3 & Part of the group’s success at achieving such a long dwell time can be credited to their choice to install backdoors on appliances within victim environments that do not support security tools, such as anti-virus or endpoint protection \\
        11 & As we detail the technical components of this attack, we can confirm that we have undertaken pre-release disclosure to the victims and provided all necessary content required to remove all known attack components from their environments \\
        15 & Both are sold as malware-as-a-service on specialized hacking forums, and both allow buyers to customize the malware with their own command and control (C\&C) and obfuscation methods \\
    \midrule
    
     & \textbf{\infolabel} \\
    \midrule
        1 & The transaction details and monetization patterns of modern eCrime reveal critical insights for organizations defending against ransomware attacks \\
        2 & EMBER BEAR appears primarily motivated to weaponize the access and data obtained during their intrusions to support information operations (IO) aimed at creating public mistrust in targeted institutions and degrading government ability to counter Russian cyber operations \\
        3 & The top referenced vulnerabilities associated with cyberattacks in H1 2022 affected Apache’s Log4J (Log4Shell), Microsoft Windows (Follina), Microsoft Exchange Server (ProxyShell), Atlassian’s Confluence, and the Java Spring Framework \\
        15 & Trellix Threat Labs includes elite researchers who have been studying cybercriminal groups and nation-backed cyber activity for years, independently and in collaboration with global government agencies \\
    \midrule
    
    & \textbf{\malwarelabel} \\
    \midrule
         1 & Industry reporting has claimed the Go-based ransomware dubbed PartyTicket (or HermeticRansom) was identified at several organizations affected by the attack,1 among other families including a sophisticated wiper CrowdStrike Intelligence tracks as DriveSlayer (HermeticWiper) \\
         2 & Once those two files have been written, one of the DLLs loaded by ProcessHacker has to be hijacked using a technique called a DLL search order hijacking \\
         3 & While it appears to primarily deliver commodity malware, Secureworks® Counter Threat Unit™ (CTU) researchers identified DarkTortilla samples delivering targeted payloads such as Cobalt Strike and Metasploit \\
         4 & We also provide a core set of detections designed to identify phishing campaign elements leveraged by both Caffeine-specific actors as well as more generalized phishing activity \\
         15 & Kraken’s presence became more apparent at the end of September, when the security researcher nao\_sec discovered that the Fallout Exploit Kit, known for delivering GandCrab ransomware, also started to deliver Kraken \\
    \midrule
    
    & \textbf{\vulnlabel} \\
    \midrule
        1 & At the start of 2022, CrowdStrike Intelligence and CrowdStrike Services investigated an incident in which PROPHET SPIDER exploited CVE-2021-22941 a remote code execution (RCE) vulnerability impacting Citrix ShareFile Storage Zones Controller â to compromise a Microsoft Internet Information Services (IIS) web server \\
        2 & Our product security (PSIRT) team collaborated with our threat intelligence team, and we quickly realized that some attempts to exploit Log4j across our customer networks manifested as illegal DNS queries \\
        3 & Duo Authentication integrates with Microsoft Windows and Active Directory (AD) to support multi-factor authentication (MFA) for both remote desktop and local logons \\
        7 & This post has been updated to reflect a new CVE (CVE-2021-45105) and associated patch (Log4j) \\
        15 & The impact of this vulnerability has the potential to be massive due to its effect on any product which has integrated the Log4j library into its applications, including products from internet giants such as Apple iCloud, Steam, Samsung Cloud storage, and thousands of additional products and services \\
    \bottomrule
    \end{tabular}
    \caption{Most influential sentences per label given by their ranking~(based on their angular similarity to the ground truth vector of that specific label)}
    \label{tab:most_important_sentences}
\end{table*}

\begin{figure}
  \includegraphics[width=\linewidth]{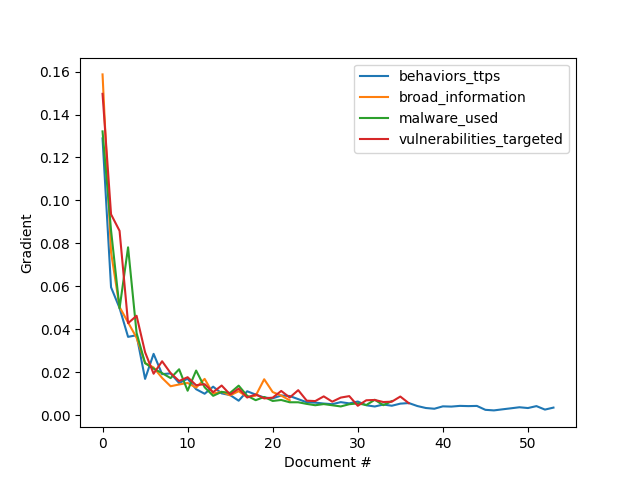}
  \caption{The ground truth gradients.}
  \label{fig:ground_truth_gradient}
\end{figure}

An excerpt of the most important sentences results for each label is shown in \autoref{tab:most_important_sentences}.
The sentences in \ttplabel{} contain a lot of general information about threat actors and their behavior~(\eg, sentence 2).
On the other hand, sentence~11 describes no behaviors or \acp{ttp} at all, which shows that it is hard to separate the labels and capture the most important parts of a document.
Moreover, documents that describe \acp{ttp} also contain information about malware or vulnerabilities, which changes the focus of the embedding.
The same can be seen at the label \infolabel.
While it clearly focuses on general information, \eg, motivation and the evolution of threat actors~(\eg, sentence 2), it shares information with the \ttplabel.
Some sentences~(\eg, sentence 3) name vulnerabilities and others~(\eg, sentence 15) the researcher-group that performed the studying of threat actors.
The label \malwarelabel{} on the other hand works well, having most sentences naming specific malware or describing how they work.
The last label, \vulnlabel{}, shows an increased occurrence of specific \acp{cve}~(\eg, sentences 1, 7), which describe vulnerabilities, and words like ``execution'' and ``bug''~(\eg, sentences 1, 2).
That shows, that while the different topics of each label can be identified by the top sentences, an overlap is noticeable.
\autoref{fig:ground_truth_gradient} shows the gradient of ground truth vectors during training, displaying the stabilization of the vectors after 10 document.

\autoref{tab:compDiffParams} shows the result of the K-fold cross-validation for 6 different sentence configurations as well as their runtime.
The adaptive amount of sentences configuration exits at 11 sentences when using 0.02 as gradient limit.
However, when comparing that configuration with the one using 50 sentences, the latter outperforms it for almost every label, showing that 0.02 as gradient limit was to high and should have been lower.
Compared to the configuration using 100 sentences on the other hand, the one with 50 sentences still performs better.
A reason is overfitting, which is confirmed by the model using all sentences, which performs even worse.

The label performing the worst over all tested configurations is \malwarelabel, with an average F1 score of 0.3, which is contrary to the clear vocabulary of this label.
It might be caused by the fact that the used malware is also mentioned in documents of other labels.

Another statistic that stands out is the fact, that the label \vulnlabel{} has a high recall, but low precision.
This might be caused by the relatively high allowed distance for this label, which could also be a influential factor for the aforementioned low values for \malwarelabel.

The very high recall and low precision for \textit{relevant} is interesting as well.
The reason for that is the calculation of the allowed distance.
The first 5 configurations use the graceful max-distance measurement as the allowed distance.
This is different in the last configuration, where the much lower, average distance of all documents per label is used as allowed distance.
It reduces the number of \emph{relevant} documents as well.
Lastly it is important to note, that the range of all values between different configurations is very high, which is a sign of a dataset that might be too small.
But while the values for the 4 labels \ttplabel, \infolabel, \malwarelabel, and \vulnlabel{} are not optimal, the metrics for the label \relevancelabel{} are still in an acceptable range with a precision of 0.77, a recall of 0.99 and a resulting F1 score of 0.86 in the best configuration.

\begin{table*}
\centering\small
\begin{tabular}{@{}rllllllllllllllllll@{}}
\toprule
\textbf{\#Sent.}          & \multicolumn{3}{c}{\textbf{10}}  & \multicolumn{3}{c}{\textbf{50}}  & \multicolumn{3}{c}{\textbf{100}}    & \multicolumn{3}{c}{\textbf{all}} & \multicolumn{3}{c}{\textbf{adaptive}} & \multicolumn{3}{c}{\textbf{average distance}} \\
\cmidrule(lr){2-4} \cmidrule(lr){5-7} \cmidrule(lr){8-10} \cmidrule(lr){11-13} \cmidrule(lr){14-16} \cmidrule(lr){17-19}
                          & Prec      & Rec       & F1       & Prec       & Rec        & F1     & Prec         & Rec          & F1    & Prec       & Rec        & F1     & Prec        & Rec         & F1        & Prec      & Rec       & F1                    \\
\ttplabels                & 0.46      & 0.64      & 0.53     & 0.7        & 0.44       & 0.5    & 0.5          & 0.22         & 0.2   & 0.46       & 0.18       & 0.17   & 0.42        & 0.67        & 0.5       & 0.81      & 0.35      & 0.46                  \\
\infolabels               & 0.45      & 0.25      & 0.32     & 0.77       & 0.48       & 0.53   & 0.73         & 0.45         & 0.52  & 0.48       & 0.28       & 0.24   & 0.4         & 0.35        & 0.36      & 0.47      & 0.45      & 0.44                  \\
\malwarelabels            & 0.65      & 0.14      & 0.22     & 0.35       & 0.39       & 0.37   & 0.29         & 0.89         & 0.35  & 0.61       & 0.28       & 0.35   & 0.27        & 0.05        & 0.08      & 0.53      & 0.34      & 0.4                   \\
\vulnlabels               & 0.23      & 0.97      & 0.37     & 0.39       & 0.91       & 0.52   & 0.91         & 0.5          & 0.64  & 0.33       & 0.92       & 0.39   & 0.39        & 0.97        & 0.49      & 0.63      & 0.52      & 0.56                  \\
\relevancelabels          & 0.61      & 0.98      & 0.76     & 0.77       & 0.99       & 0.86   & 0.66         & 0.97         & 0.78  & 0.62       & 0.99       & 0.76   & 0.62        & 0.99        & 0.76      & 0.97      & 0.62      & 0.75                  \\
\cmidrule(lr){2-4} \cmidrule(lr){5-7} \cmidrule(lr){8-10} \cmidrule(lr){11-13} \cmidrule(lr){14-16} \cmidrule(lr){17-19}
Runtime                   & \multicolumn{3}{c}{10m 34s}      & \multicolumn{3}{c}{46m 1s}       & \multicolumn{3}{c}{1h 1m 55s}       & \multicolumn{3}{c}{2h 49m 21s}   & \multicolumn{3}{c}{2h 48m 18s}        & \multicolumn{3}{c}{2h 17m 59s}                \\
\bottomrule
\end{tabular}
\caption{Comparison of Different Parameters. The abbreviations \ttplabels, \infolabels, \malwarelabels, \vulnlabels, and \relevancelabels{} stand for the labels \ttplabel, \infolabel, \malwarelabel, \vulnlabel, and overall \relevancelabel, respectively.}
\label{tab:compDiffParams}
\end{table*}

In \autoref{tab:comparison_of_bert_models} the comparison of different models is shown.
Surprisingly, while the fine-tuned CySecBERT gets similar values as the \ac{sbert} model, the more general \ac{bert} model outperforms both models.
This could be a sign that CySecBERT might be too specialized, while the task of classifying online content requires a more generalized model.

\begin{table}
\centering\small
\begin{tabular}{@{}rllllllllllll@{}}
\toprule
                 & \multicolumn{3}{c}{\textbf{CySecBERT}~\citep{bayer_cysecbert_2024}} &  \multicolumn{3}{c}{\textbf{BERT}~\citep{devlin_bert_2018}} & \multicolumn{3}{c}{\textbf{SBERT}~\citep{reimers_sentencebert_2019}} \\
\cmidrule(lr){2-4} \cmidrule(lr){5-7} \cmidrule(lr){8-10}
                 & Prec       & Rec         & F1          &  Prec        & Rec        & F1     & Prec       & Rec       & F1        \\
\ttplabels       & 0.42       & 0.67        & 0.50        &  0.42        & 0.67       & 0.51   & 0.52       & 0.31      & 0.32      \\
\infolabels      & 0.4        & 0.35        & 0.36        &  0.45        & 0.55       & 0.47   & 0.79       & 0.45      & 0.54      \\
\malwarelabels   & 0.27       & 0.05        & 0.08        &  0.43        & 0.17       & 0.22   & 0.39       & 0.16      & 0.22      \\
\vulnlabels      & 0.39       & 0.97        & 0.49        &  0.41        & 0.91       & 0.56   & 0.33       & 0.92      & 0.4       \\
\relevancelabels & 0.62       & 0.99        & 0.76        &  0.77        & 0.98       & 0.86   & 0.73       & 0.98      & 0.83      \\
\bottomrule
\end{tabular}
\caption{Comparison of Different \ac*{bert} Models}
\label{tab:comparison_of_bert_models}
\end{table}

\subsection{Evaluation of \threatcrawl{}}
\label{subsec:eval_crawl_bot}
Using the results from the previous evaluation, the model \ac{bert} with 50 sentences was used to evaluate the overall performance of \threatcrawl{}.
To calculate the allowed distance, the maximum distance per label was used.
This leads to more false positives, but also reduces the false negatives, which is considered the better trade-off for this usage.

To test \threatcrawl{}, \num{1000} \acp{url} were crawled, with a runtime of 19 minutes and 15 seconds.
Out of these \num{1000} \acp{url}, \num{512} were classified as relevant, which results in harvest rate of relevant documents of 51.2\%.
This is a huge improvement compared to \citeauthor{koloveas_crawler_2019}'s crawler~\citep{koloveas_crawler_2019} with a harvest rate of 1\%.
When looking at the harvest rate, it is important to keep in mind that the \ac{ml} model with more false positives was used and the fact, that the crawler changes its path dynamically, not following routes with non-relevant documents.

\begin{table}
    \centering
    \begin{tabular}{@{} rll @{}}
    \toprule
    & \acs*{url} & Label \\
    \midrule
        1  & \href{https://www.fireeye.com/blog/threat-research/2017/04/fin7-phishing-lnk.html}{www.fireeye.com/blog/\dots} & \ttplabels \\
        2  & \href{https://www.zscaler.com/blogs/security-research/return-higaisa-apt}{www.zscaler.com/blogs/\dots} & \ttplabels \\
        3  & \href{https://blog.talosintelligence.com/page/2/}{blog.talosintelligence.com/\dots} & \vulnlabels \\
        4  & \href{https://blog.talosintelligence.com/page/5/}{blog.talosintelligence.com/\dots} & \vulnlabels \\
        5  & \href{https://blog.talosintelligence.com/author/joe-marshall/}{blog.talosintelligence.com/\dots} & \ttplabels \\
        6  & \href{https://blog.talosintelligence.com/category/threat-advisory/}{blog.talosintelligence.com\dots} & \vulnlabels \\
        7  & \href{https://www.proofpoint.com/us/threat-insight/post/lookback-malware-targets-united-states-utilities-sector-phishing-attacks}{www.proofpoint.com/\dots} & \ttplabels \\
        8  & \href{https://research.nccgroup.com/2018/03/10/apt15-is-alive-and-strong-an-analysis-of-royalcli-and-royaldns/}{research.nccgroup.com/\dots} & \vulnlabels \\
        9  & \href{https://www.proofpoint.com/us/threat-insight/post/APT-targets-russia-belarus-zerot-plugx}{www.proofpoint.com/\dots} & \ttplabels \\
        \addlinespace
        & \dots \\
        \addlinespace
        507  & \href{https://docs.github.com/en/enterprise-cloud@latest/admin/user-management/managing-organizations-in-your-enterprise/removing-organizations-from-your-enterprise}{docs.github.com/\dots} & \vulnlabels \\
        508  & \href{https://support.mozilla.org/en-US/questions/new/firefox-private-network-vpn}{support.mozilla.org/\dots} & \vulnlabels \\
        509  & \href{https://www.mozilla.org/privacy/websites/\#cookies}{support.mozilla.org/\dots} & \vulnlabels \\
        510  & \href{https://www.mozilla.org/privacy/websites/\#data-tools}{support.mozilla.org/\dots} & \vulnlabels \\
        511  & \href{https://attack.mitre.org/software/S1028/}{attack.mitre.org/\dots} & \vulnlabels \\
    \bottomrule
    \end{tabular}
    \caption{Sorted \threatcrawl{} results ranked by their relevance with the according label.}
    \label{tab:sorted_crawler_results}
\end{table}

\autoref{tab:sorted_crawler_results} shows an excerpt of the top entries of the list of relevant \acp{url} and their label given by \threatcrawl{} sorted by the lowest relative distance~\see{subsubsec:classify_single_documents}.
Except number two, three, and four, all of the shown top documents are blog posts or news articles.
The lower ranked \acp{url} are somewhat related to the \ac{cti} domain and describe content like vulnerabilities or \acp{apt}, but they do not contain the specific type of content that is considered relevant for this work.
This shows, that the usage of the relative distance as an effective way to rank the downloaded documents.

Overview pages of news sites or blogs that contain a list of all recent articles do not get consistently classified as relevant.
This can lead to situations, where documents can not get reached by \threatcrawl{} because it stops one hop before those pages.
The reason for this behavior is the fact, that overview pages do not contain a lot of information.
When extracting the main content of those pages, not many sentences get extracted which leads to very unstable embeddings.

\begin{figure}
    \centering
    \includegraphics[width=\linewidth]{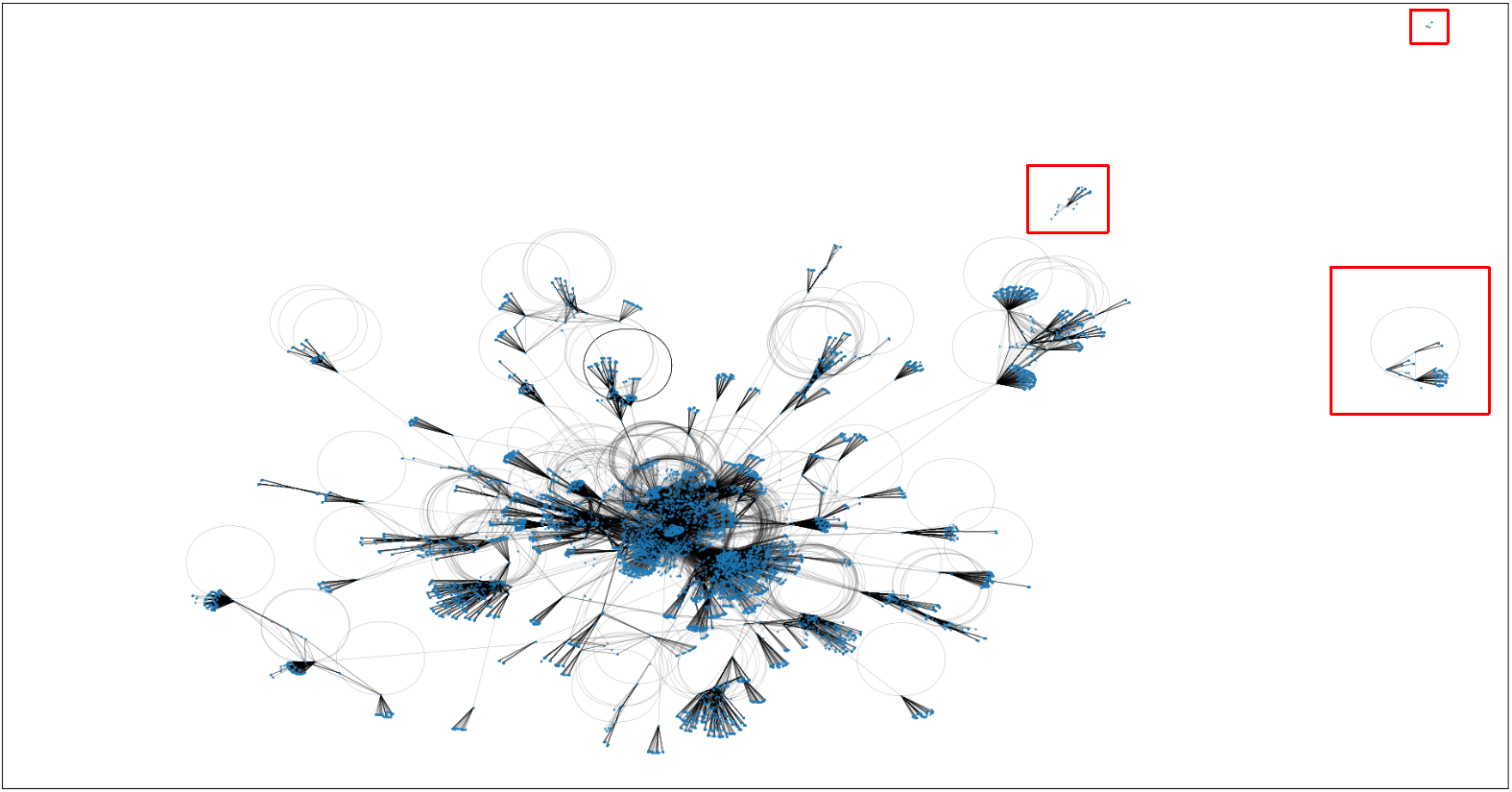}
    \caption{Multigraph of the crawled webpages.}
    \label{fig:full_url_map}
\end{figure}

Multigraphs can be used to visualize the structure of all downloaded documents and their relationship to each other.
In this graph, nodes represent documents and an edge between a document A and a document B represents a reference to B that appears in A.
A multigraph of all explored documents is depicted in \autoref{fig:full_url_map}.
As described in \autoref{subsubsec:choosing_right_seed}, choosing the right seed is important to avoid getting stuck in isolated groups of documents.
Even though a fixed set of pre-chosen \acp{url} was used as a seed, the multigraph of all documents shows, that most documents can be reached over the available edges.
Only three small isolated groups can be seen in the top right~(marked red).
This shows, that the seed is sufficient enough for this task.
It is important to note, that this highly depends on the nature of the examined domain.
News articles and blog entries in the \acp{cti} domain are highly interconnected, which severely lowers the appearance of isolated groups of content.
This does not have to be the case for content of other domains.

\section{Conclusion}
\label{sec:conclusion}
This section concludes the paper by outlining our contributions, answering our research questions~\see{sec:introduction}, presenting limitations, implications, and future work.

\paragraph{Contributions}
\threatcrawl{} uses a new approach to find and rank \ac{cti}-relevant information by downloading and classifying documents before references are extracted, which answers our main research question~\rq{}.

It answers \textbf{RQ1} on how websites in \ac{cti} can be efficiently crawled and ranked~\see{sec:introduction} by using dynamic pathing, where non-relevant documents are ignored to avoid whole clusters of non-relevant documents.
This leading to a harvest rate of 52\% for relevant documents.
Other approaches use a two-step approach by downloading huge amounts of content first and filter non-relevant documents out later.
Contrary to that, the design presented manages to perform both tasks at the same time.
Thus, lots of resources like disk space, network resources and server capacities can be saved.

\textbf{RQ2} \emph{``How can websites be categorized automatically as \ac{cti}-relevant?''} is answered by using contextualized document embeddings.
Those are calculated over the training set by averaging generated contextualized document embeddings for each document in the set.
Document embeddings are generated by averaging sentence embeddings which are generated by averaging contextualized embeddings for each token in a sentence.
During the crawl process, the document is accordingly labeled.
The evaluation in \autoref{sec:evaluation} shows good results for the implemented models.
Surprisingly, the general \ac{bert} model~\citep{devlin_bert_2018} outperforms the more specialized CySecBERT~\citep{bayer_cysecbert_2024} and \ac{sbert} model~\citep{reimers_sentencebert_2019}.

The generated ground truth vectors can be seen as the most \emph{relevant} documents.
The closer the embedding of a document, the more \emph{relevant} it is.
Therefore, documents can be sorted based on their distance from the according ground truth vector.
To make those distances comparable over all labels, those distances get converted into relative distances.
Hence, the final list is a list of all relevant documents sorted by their relative distance from their ground truth vector, answering \textbf{RQ3} on how the crawling and and labeling process can be combined.

\paragraph{Implications}
\threatcrawl{} can be trained with a labeled dataset and then started to crawl any amount of \acp{url}.
Due to its flexible structure, every module can be changed and optimized on its own.
The used one-step approach, its dynamic pathing and the implemented embedding and classification approach can be used to refine the search for \ac{cti}-relevant content as well as foundation for other focused crawler and the search for content in other domains.

\paragraph{Limitations and Future Work}
Unfortunately, some limitations occurred during the development of \threatcrawl{}.
The high range of metrics in the evaluation~\see{sec:evaluation} showed, that the created dataset is too sparse and the labels chosen overlap.
Future work could therefore start to define more specific labels and a build bigger dataset to generate better performing models.
Another limitation of this work is the small seed.
As mentioned by \citet{vieira_finding_2016}, choosing the right seed is essential for focused crawlers.
Methods used by \citet{zhang_dsdd_2021} might show further improvements in this regard.
And even though \autoref{fig:full_url_map} shows very few isolated components, an optimized seed could reveal even more unique parts of the \ac{cti}-domain.
One last proposal for future work is to implement solutions to avoid the problem of overview pages as described in \autoref{subsec:eval_crawl_bot}.
This could be for example done by detecting those pages using other \ac{nlp} techniques or by implementing a two-hop approach, that checks references of non-relevant documents for one more hop before stopping to follow them.
That way, the amount of occurrences where one non-relevant page prevents the access to multiple potential relevant pages can be minimized.

\begin{acks}
This work was supported by the German Federal Ministry for Education and Research~(BMBF) in the project CYWARN~(13N15407)
and
German Federal Ministry of Education and Research and the Hessian Ministry of Higher Education, Research, Science and the Arts within their joint support of the National Research Center for Applied Cybersecurity ATHENE.
\end{acks}

\bibliographystyle{ACM-Reference-Format}
\bibliography{threatcrawl}


\begin{thebibliography}{37}


\ifx \showCODEN    \undefined \def \showCODEN     #1{\unskip}     \fi
\ifx \showDOI      \undefined \def \showDOI       #1{#1}\fi
\ifx \showISBNx    \undefined \def \showISBNx     #1{\unskip}     \fi
\ifx \showISBNxiii \undefined \def \showISBNxiii  #1{\unskip}     \fi
\ifx \showISSN     \undefined \def \showISSN      #1{\unskip}     \fi
\ifx \showLCCN     \undefined \def \showLCCN      #1{\unskip}     \fi
\ifx \shownote     \undefined \def \shownote      #1{#1}          \fi
\ifx \showarticletitle \undefined \def \showarticletitle #1{#1}   \fi
\ifx \showURL      \undefined \def \showURL       {\relax}        \fi
\providecommand\bibfield[2]{#2}
\providecommand\bibinfo[2]{#2}
\providecommand\natexlab[1]{#1}
\providecommand\showeprint[2][]{arXiv:#2}

\bibitem[Barbaresi(2021)]%
        {barbaresi_trafilatura_2021}
\bibfield{author}{\bibinfo{person}{Adrien Barbaresi}.}
  \bibinfo{year}{2021}\natexlab{}.
\newblock \showarticletitle{Trafilatura: {{A Web Scraping Library}} and
  {{Command-Line Tool}} for {{Text Discovery}} and {{Extraction}}}. In
  \bibinfo{booktitle}{\emph{{{ACL}}/{{IJCNLP}}'21}}. \bibinfo{publisher}{ACL},
  \bibinfo{pages}{122--131}.
\newblock
\urldef\tempurl%
\url{https://doi.org/10.18653/v1/2021.acl-demo.15}
\showDOI{\tempurl}


\bibitem[Bayer et~al\mbox{.}(2024)]%
        {bayer_cysecbert_2024}
\bibfield{author}{\bibinfo{person}{Markus Bayer}, \bibinfo{person}{Philipp
  Kuehn}, \bibinfo{person}{Ramin Shanehsaz}, {and} \bibinfo{person}{Christian
  Reuter}.} \bibinfo{year}{2024}\natexlab{}.
\newblock \showarticletitle{{{CySecBERT}}: {{A Domain-Adapted Language Model}}
  for the {{Cybersecurity Domain}}}.
\newblock \bibinfo{journal}{\emph{ACM Transactions on Privacy and Security}}
  (\bibinfo{year}{2024}).
\newblock
\urldef\tempurl%
\url{https://doi.org/10.1145/3652594}
\showDOI{\tempurl}


\bibitem[Bianco(2014)]%
        {bianco_pyramid_2014}
\bibfield{author}{\bibinfo{person}{David~J. Bianco}.}
  \bibinfo{year}{2014}\natexlab{}.
\newblock \bibinfo{title}{The {{Pyramid}} of {{Pain}}}.
\newblock
\newblock


\bibitem[Brown and Lee(2021)]%
        {brown_sans_2021}
\bibfield{author}{\bibinfo{person}{Rebekah Brown} {and}
  \bibinfo{person}{Robert~M. Lee}.} \bibinfo{year}{2021}\natexlab{}.
\newblock \bibinfo{booktitle}{\emph{{{SANS Cyber Threat Intelligence Survey}}
  2021}}.
\newblock \bibinfo{type}{{T}echnical {R}eport}. \bibinfo{institution}{SANS}.
\newblock


\bibitem[Brown and Stirparo(2022)]%
        {brown_sans_2022}
\bibfield{author}{\bibinfo{person}{Rebekah Brown} {and}
  \bibinfo{person}{Pasquale Stirparo}.} \bibinfo{year}{2022}\natexlab{}.
\newblock \bibinfo{booktitle}{\emph{{{SANS Cyber Threat Intelligence Survey}}
  2022}}.
\newblock \bibinfo{type}{{T}echnical {R}eport}. \bibinfo{institution}{SANS}.
\newblock


\bibitem[Chakrabarti et~al\mbox{.}(1999)]%
        {chakrabarti_focused_1999}
\bibfield{author}{\bibinfo{person}{Soumen Chakrabarti}, \bibinfo{person}{Martin
  {van den Berg}}, {and} \bibinfo{person}{Byron Dom}.}
  \bibinfo{year}{1999}\natexlab{}.
\newblock \showarticletitle{Focused crawling: {{A}} new approach to
  topic-specific {{Web}} resource discovery}.
\newblock \bibinfo{journal}{\emph{Computer Networks}}  \bibinfo{volume}{31}
  (\bibinfo{year}{1999}), \bibinfo{pages}{1623--1640}.
\newblock
\urldef\tempurl%
\url{https://doi.org/10.1016/S1389-1286(99)00052-3}
\showDOI{\tempurl}


\bibitem[{Chris Sanders} and {Jason Smith}(2013)]%
        {chrissanders_applied_2013}
\bibfield{author}{\bibinfo{person}{{Chris Sanders}} {and}
  \bibinfo{person}{{Jason Smith}}.} \bibinfo{year}{2013}\natexlab{}.
\newblock \bibinfo{booktitle}{\emph{Applied {{Network Security Monitoring}}}
  (\bibinfo{edition}{1} ed.)}.
\newblock \bibinfo{publisher}{Elsevier}.
\newblock


\bibitem[Denker et~al\mbox{.}(2019)]%
        {denker_darknets_2019}
\bibfield{author}{\bibinfo{person}{Kai Denker}, \bibinfo{person}{Marcel
  Sch{\"a}fer}, {and} \bibinfo{person}{Martin Steinebach}.}
  \bibinfo{year}{2019}\natexlab{}.
\newblock \showarticletitle{Darknets as {{Tools}} for {{Cyber Warfare}}}.
\newblock In \bibinfo{booktitle}{\emph{Information {{Technology}} for {{Peace}}
  and {{Security}}: {{IT Applications}} and {{Infrastructures}} in
  {{Conflicts}}, {{Crises}}, {{War}}, and {{Peace}}}},
  \bibfield{editor}{\bibinfo{person}{Christian Reuter}} (Ed.).
  \bibinfo{publisher}{Springer Fachmedien}, \bibinfo{pages}{107--135}.
\newblock
\urldef\tempurl%
\url{https://doi.org/10.1007/978-3-658-25652-4_6}
\showDOI{\tempurl}


\bibitem[Devlin et~al\mbox{.}(2018)]%
        {devlin_bert_2018}
\bibfield{author}{\bibinfo{person}{Jacob Devlin}, \bibinfo{person}{Ming-Wei
  Chang}, \bibinfo{person}{Kenton Lee}, {and} \bibinfo{person}{Kristina
  Toutanova}.} \bibinfo{year}{2018}\natexlab{}.
\newblock \showarticletitle{{{BERT}}: {{Pre-training}} of {{Deep Bidirectional
  Transformers}} for {{Language Understanding}}}.
\newblock \bibinfo{journal}{\emph{NORD'19}}  \bibinfo{volume}{1}
  (\bibinfo{year}{2018}), \bibinfo{pages}{4171--4186}.
\newblock
\urldef\tempurl%
\url{https://doi.org/10.18653/v1/N19-1423}
\showDOI{\tempurl}


\bibitem[Elbaz et~al\mbox{.}(2020)]%
        {elbaz_fighting_2020}
\bibfield{author}{\bibinfo{person}{Cl{\'e}ment Elbaz}, \bibinfo{person}{Louis
  Rilling}, {and} \bibinfo{person}{Christine Morin}.}
  \bibinfo{year}{2020}\natexlab{}.
\newblock \showarticletitle{Fighting {{N-day}} vulnerabilities with automated
  {{CVSS}} vector prediction at disclosure}. In
  \bibinfo{booktitle}{\emph{Proceedings of the 15th {{International
  Conference}} on {{Availability}}, {{Reliability}} and {{Security}}}}.
  \bibinfo{publisher}{ACM}, \bibinfo{pages}{1--10}.
\newblock
\urldef\tempurl%
\url{https://doi.org/10.1145/3407023.3407038}
\showDOI{\tempurl}


\bibitem[Hastie et~al\mbox{.}(2009)]%
        {hastie_elements_2009}
\bibfield{author}{\bibinfo{person}{Trevor Hastie}, \bibinfo{person}{Robert
  Tibshirani}, {and} \bibinfo{person}{Jerome Friedman}.}
  \bibinfo{year}{2009}\natexlab{}.
\newblock \bibinfo{booktitle}{\emph{The {{Elements}} of {{Statistical
  Learning}}}}.
\newblock \bibinfo{publisher}{Springer}.
\newblock
\urldef\tempurl%
\url{https://doi.org/10.1007/978-0-387-84858-7}
\showDOI{\tempurl}


\bibitem[Hinchy(2022)]%
        {hinchy_voice_2022}
\bibfield{author}{\bibinfo{person}{Eoin Hinchy}.}
  \bibinfo{year}{2022}\natexlab{}.
\newblock \bibinfo{booktitle}{\emph{Voice of the {{SOC Analyst}}}}.
\newblock \bibinfo{type}{{T}echnical {R}eport}. \bibinfo{institution}{Tines}.
  \bibinfo{pages}{39} pages.
\newblock


\bibitem[Husari et~al\mbox{.}(2017)]%
        {husari_ttpdrill_2017}
\bibfield{author}{\bibinfo{person}{Ghaith Husari}, \bibinfo{person}{Ehab
  {Al-Shaer}}, \bibinfo{person}{Mohiuddin Ahmed}, \bibinfo{person}{Bill Chu},
  {and} \bibinfo{person}{Xi Niu}.} \bibinfo{year}{2017}\natexlab{}.
\newblock \showarticletitle{{{TTPDrill}}: {{Automatic}} and {{Accurate
  Extraction}} of {{Threat Actions}} from {{Unstructured Text}} of {{CTI
  Sources}}}. In \bibinfo{booktitle}{\emph{Proceedings of the 33rd {{Annual
  Computer Security Applications Conference}}}}.
  \bibinfo{publisher}{Association for Computing Machinery},
  \bibinfo{pages}{103--115}.
\newblock
\urldef\tempurl%
\url{https://doi.org/10.1145/3134600.3134646}
\showDOI{\tempurl}


\bibitem[Koloveas et~al\mbox{.}(2021)]%
        {koloveas_intime_2021}
\bibfield{author}{\bibinfo{person}{Paris Koloveas}, \bibinfo{person}{Thanasis
  Chantzios}, \bibinfo{person}{Sofia Alevizopoulou}, \bibinfo{person}{Spiros
  Skiadopoulos}, {and} \bibinfo{person}{Christos Tryfonopoulos}.}
  \bibinfo{year}{2021}\natexlab{}.
\newblock \showarticletitle{{{inTIME}}: {{A Machine Learning-Based Framework}}
  for {{Gathering}} and {{Leveraging Web Data}} to {{Cyber-Threat
  Intelligence}}}.
\newblock \bibinfo{journal}{\emph{Electronicsweek}}  \bibinfo{volume}{10}
  (\bibinfo{year}{2021}), \bibinfo{pages}{818}.
\newblock
\urldef\tempurl%
\url{https://doi.org/10.3390/electronics10070818}
\showDOI{\tempurl}


\bibitem[Koloveas et~al\mbox{.}(2019)]%
        {koloveas_crawler_2019}
\bibfield{author}{\bibinfo{person}{Paris Koloveas}, \bibinfo{person}{Thanasis
  Chantzios}, \bibinfo{person}{Christos Tryfonopoulos}, {and}
  \bibinfo{person}{Spiros Skiadopoulos}.} \bibinfo{year}{2019}\natexlab{}.
\newblock \showarticletitle{A {{Crawler Architecture}} for {{Harvesting}} the
  {{Clear}}, {{Social}}, and {{Dark Web}} for {{IoT-Related Cyber-Threat
  Intelligence}}}. In \bibinfo{booktitle}{\emph{2019 {{IEEE World Congress}} on
  {{Services}} ({{SERVICES}})}}. \bibinfo{publisher}{IEEE},
  \bibinfo{pages}{3--8}.
\newblock
\urldef\tempurl%
\url{https://doi.org/10.1109/SERVICES.2019.00016}
\showDOI{\tempurl}


\bibitem[Krotov et~al\mbox{.}(2020)]%
        {krotov_legality_2020}
\bibfield{author}{\bibinfo{person}{Vlad Krotov}, \bibinfo{person}{Leigh
  Johnson}, {and} \bibinfo{person}{Leiser Silva}.}
  \bibinfo{year}{2020}\natexlab{}.
\newblock \showarticletitle{Legality and {{Ethics}} of {{Web Scraping}}}.
\newblock \bibinfo{journal}{\emph{Communications of the Association for
  Information Systems}}  \bibinfo{volume}{47} (\bibinfo{year}{2020}),
  \bibinfo{pages}{539--563}.
\newblock
\urldef\tempurl%
\url{https://doi.org/10.17705/1CAIS.04724}
\showDOI{\tempurl}


\bibitem[Kuehn et~al\mbox{.}(2022)]%
        {kuehn_notion_2022}
\bibfield{author}{\bibinfo{person}{Philipp Kuehn}, \bibinfo{person}{Julian
  B{\"a}umler}, \bibinfo{person}{Marc-Andr{\'e} Kaufhold},
  \bibinfo{person}{Marc Wendelborn}, {and} \bibinfo{person}{Christian Reuter}.}
  \bibinfo{year}{2022}\natexlab{}.
\newblock \showarticletitle{The {{Notion}} of {{Relevance}} in
  {{Cybersecurity}}: {{A Categorization}} of {{Security Tools}} and
  {{Deduction}} of {{Relevance Notions}}}. In
  \bibinfo{booktitle}{\emph{Workshop-{{Proceedings Mensch}} und {{Computer}}}}.
  \bibinfo{publisher}{Gesellschaft f{\"u}r Informatik}.
\newblock
\urldef\tempurl%
\url{https://doi.org/10.18420/muc2022-mci-ws01-220}
\showDOI{\tempurl}


\bibitem[Kuehn et~al\mbox{.}(2021)]%
        {kuehn_ovana_2021}
\bibfield{author}{\bibinfo{person}{Philipp Kuehn}, \bibinfo{person}{Markus
  Bayer}, \bibinfo{person}{Marc Wendelborn}, {and} \bibinfo{person}{Christian
  Reuter}.} \bibinfo{year}{2021}\natexlab{}.
\newblock \showarticletitle{{{OVANA}}: {{An Approach}} to {{Analyze}} and
  {{Improve}} the {{Information Quality}} of {{Vulnerability Databases}}}. In
  \bibinfo{booktitle}{\emph{{{ARES}} '21: {{Proceedings}} of the 16th
  {{International Conference}} on {{Availability}}, {{Reliability}} and
  {{Security}}}}. \bibinfo{publisher}{ACM}, \bibinfo{pages}{11}.
\newblock
\urldef\tempurl%
\url{https://doi.org/10.1145/3465481.3465744}
\showDOI{\tempurl}


\bibitem[Le~Sceller et~al\mbox{.}(2017)]%
        {lesceller_sonar_2017}
\bibfield{author}{\bibinfo{person}{Quentin Le~Sceller},
  \bibinfo{person}{ElMouatez~Billah Karbab}, \bibinfo{person}{Mourad Debbabi},
  {and} \bibinfo{person}{Farkhund Iqbal}.} \bibinfo{year}{2017}\natexlab{}.
\newblock \showarticletitle{{{SONAR}}: {{Automatic Detection}} of {{Cyber
  Security Events}} over the {{Twitter Stream}}}. In
  \bibinfo{booktitle}{\emph{{{ARES}}'17}}. \bibinfo{publisher}{ACM},
  \bibinfo{pages}{1--11}.
\newblock
\urldef\tempurl%
\url{https://doi.org/10.1145/3098954.3098992}
\showDOI{\tempurl}


\bibitem[Lee(2020)]%
        {lee_2020_2020}
\bibfield{author}{\bibinfo{person}{Robert~M Lee}.}
  \bibinfo{year}{2020}\natexlab{}.
\newblock \showarticletitle{2020 {{SANS}} cyber threat intelligence ({{CTI}})
  survey}.
\newblock  (\bibinfo{year}{2020}), \bibinfo{pages}{17}.
\newblock


\bibitem[Liao et~al\mbox{.}(2016)]%
        {liao_acing_2016}
\bibfield{author}{\bibinfo{person}{Xiaojing Liao}, \bibinfo{person}{Kan Yuan},
  \bibinfo{person}{Xiaofeng Wang}, \bibinfo{person}{Zhou Li},
  \bibinfo{person}{Luyi Xing}, {and} \bibinfo{person}{Raheem Beyah}.}
  \bibinfo{year}{2016}\natexlab{}.
\newblock \showarticletitle{Acing the {{IOC}} game: {{Toward}} automatic
  discovery and analysis of open-source cyber threat intelligence}. In
  \bibinfo{booktitle}{\emph{{{CSS}}'16}}, Vol.~\bibinfo{volume}{24-28-Octo}.
  \bibinfo{publisher}{ACM Press}, \bibinfo{pages}{755--766}.
\newblock
\urldef\tempurl%
\url{https://doi.org/10.1145/2976749.2978315}
\showDOI{\tempurl}


\bibitem[Liu et~al\mbox{.}(2023)]%
        {liu_focused_2023}
\bibfield{author}{\bibinfo{person}{Wenjun Liu}, \bibinfo{person}{Yu He},
  \bibinfo{person}{Jing Wu}, \bibinfo{person}{Yajun Du}, \bibinfo{person}{Xing
  Liu}, \bibinfo{person}{Tiejun Xi}, \bibinfo{person}{Zurui Gan},
  \bibinfo{person}{Pengjun Jiang}, {and} \bibinfo{person}{Xiaoping Huang}.}
  \bibinfo{year}{2023}\natexlab{}.
\newblock \showarticletitle{A focused crawler based on semantic disambiguation
  vector space model}.
\newblock \bibinfo{journal}{\emph{Complex \& Intelligent Systems}}
  \bibinfo{volume}{9} (\bibinfo{year}{2023}), \bibinfo{pages}{345--366}.
\newblock
\urldef\tempurl%
\url{https://doi.org/10.1007/s40747-022-00707-8}
\showDOI{\tempurl}


\bibitem[Mikolov et~al\mbox{.}(2013)]%
        {mikolov_efficient_2013}
\bibfield{author}{\bibinfo{person}{Tomas Mikolov}, \bibinfo{person}{Kai Chen},
  \bibinfo{person}{G. Corrado}, {and} \bibinfo{person}{J. Dean}.}
  \bibinfo{year}{2013}\natexlab{}.
\newblock \showarticletitle{Efficient {{Estimation}} of {{Word
  Representations}} in {{Vector Space}}}. In
  \bibinfo{booktitle}{\emph{Proceedings of the {{International Conference}} on
  {{Learning Representations}}}}.
\newblock


\bibitem[Peters et~al\mbox{.}(2018)]%
        {peters_deep_2018}
\bibfield{author}{\bibinfo{person}{Matthew~E. Peters}, \bibinfo{person}{Mark
  Neumann}, \bibinfo{person}{Mohit Iyyer}, \bibinfo{person}{Matt Gardner},
  \bibinfo{person}{Christopher Clark}, \bibinfo{person}{Kenton Lee}, {and}
  \bibinfo{person}{Luke Zettlemoyer}.} \bibinfo{year}{2018}\natexlab{}.
\newblock \showarticletitle{Deep {{Contextualized Word Representations}}}. In
  \bibinfo{booktitle}{\emph{{{NAACL}}'18}}. \bibinfo{publisher}{Association for
  Computational Linguistics}, \bibinfo{pages}{2227--2237}.
\newblock
\urldef\tempurl%
\url{https://doi.org/10.18653/v1/N18-1202}
\showDOI{\tempurl}


\bibitem[Preuveneers and Joosen(2022)]%
        {preuveneers_privacypreserving_2022}
\bibfield{author}{\bibinfo{person}{Davy Preuveneers} {and}
  \bibinfo{person}{Wouter Joosen}.} \bibinfo{year}{2022}\natexlab{}.
\newblock \showarticletitle{Privacy-{{Preserving Polyglot Sharing}} and
  {{Analysis}} of {{Confidential Cyber Threat Intelligence}}}. In
  \bibinfo{booktitle}{\emph{Proceedings of the 17th {{International
  Conference}} on {{Availability}}, {{Reliability}} and {{Security}}}}.
  \bibinfo{publisher}{Association for Computing Machinery},
  \bibinfo{pages}{1--11}.
\newblock
\urldef\tempurl%
\url{https://doi.org/10.1145/3538969.3538982}
\showDOI{\tempurl}


\bibitem[Ramnani et~al\mbox{.}(2017)]%
        {ramnani_semiautomated_2017}
\bibfield{author}{\bibinfo{person}{Roshni~R. Ramnani}, \bibinfo{person}{Karthik
  Shivaram}, \bibinfo{person}{Shubhashis Sengupta}, {and}
  \bibinfo{person}{Annervaz~K. M.}} \bibinfo{year}{2017}\natexlab{}.
\newblock \showarticletitle{Semi-{{Automated Information Extraction}} from
  {{Unstructured Threat Advisories}}}. In \bibinfo{booktitle}{\emph{Proceedings
  of the 10th {{Innovations}} in {{Software Engineering Conference}}}}.
  \bibinfo{publisher}{Association for Computing Machinery},
  \bibinfo{pages}{181--187}.
\newblock
\urldef\tempurl%
\url{https://doi.org/10.1145/3021460.3021482}
\showDOI{\tempurl}


\bibitem[{\v R}eh{\r u}{\v r}ek and Sojka(2010)]%
        {rehurek_software_2010}
\bibfield{author}{\bibinfo{person}{Radim {\v R}eh{\r u}{\v r}ek} {and}
  \bibinfo{person}{Petr Sojka}.} \bibinfo{year}{2010}\natexlab{}.
\newblock \showarticletitle{Software {{Framework}} for {{Topic Modelling}} with
  {{Large Corpora}}}. \bibinfo{pages}{45--50}.
\newblock
\urldef\tempurl%
\url{https://doi.org/10.13140/2.1.2393.1847}
\showDOI{\tempurl}


\bibitem[Reimers and Gurevych(2019)]%
        {reimers_sentencebert_2019}
\bibfield{author}{\bibinfo{person}{Nils Reimers} {and} \bibinfo{person}{Iryna
  Gurevych}.} \bibinfo{year}{2019}\natexlab{}.
\newblock \showarticletitle{Sentence-{{BERT}}: {{Sentence Embeddings}} using
  {{Siamese BERT-Networks}}}. In \bibinfo{booktitle}{\emph{Proceedings of the
  2019 {{Conference}} on {{Empirical Methods}} in {{Natural Language
  Processing}} and the 9th {{International Joint Conference}} on {{Natural
  Language Processing}} ({{EMNLP-IJCNLP}})}},
  \bibfield{editor}{\bibinfo{person}{Kentaro Inui}, \bibinfo{person}{Jing
  Jiang}, \bibinfo{person}{Vincent Ng}, {and} \bibinfo{person}{Xiaojun Wan}}
  (Eds.). \bibinfo{publisher}{Association for Computational Linguistics},
  \bibinfo{pages}{3982--3992}.
\newblock
\urldef\tempurl%
\url{https://doi.org/10.18653/v1/D19-1410}
\showDOI{\tempurl}


\bibitem[Riebe et~al\mbox{.}(2021)]%
        {riebe_cysecalert_2021}
\bibfield{author}{\bibinfo{person}{Thea Riebe}, \bibinfo{person}{Tristan
  Wirth}, \bibinfo{person}{Markus Bayer}, \bibinfo{person}{Philipp Kuehn},
  \bibinfo{person}{Marc-Andr{\'e} Kaufhold}, \bibinfo{person}{Volker Knauthe},
  \bibinfo{person}{Stefan Guthe}, {and} \bibinfo{person}{Christian Reuter}.}
  \bibinfo{year}{2021}\natexlab{}.
\newblock \showarticletitle{{{CySecAlert}}: {{An Alert Generation System}} for
  {{Cyber Security Events Using Open Source Intelligence Data}}}. In
  \bibinfo{booktitle}{\emph{Information and {{Communications Security}}
  ({{ICICS}})}}. \bibinfo{pages}{429--446}.
\newblock
\urldef\tempurl%
\url{https://doi.org/10.1007/978-3-030-86890-1_24}
\showDOI{\tempurl}


\bibitem[Shackleford and Northcutt(2015)]%
        {shackleford_whos_2015}
\bibfield{author}{\bibinfo{person}{Dave Shackleford} {and}
  \bibinfo{person}{Stephen Northcutt}.} \bibinfo{year}{2015}\natexlab{}.
\newblock \bibinfo{booktitle}{\emph{Who's using cyberthreat intelligence and
  how}}.
\newblock \bibinfo{type}{{T}echnical {R}eport}. \bibinfo{pages}{24} pages.
\newblock


\bibitem[Shin et~al\mbox{.}(2021)]%
        {shin_twiti_2021}
\bibfield{author}{\bibinfo{person}{Hyejin Shin}, \bibinfo{person}{WooChul
  Shim}, \bibinfo{person}{Saebom Kim}, \bibinfo{person}{Sol Lee},
  \bibinfo{person}{Yong~Goo Kang}, {and} \bibinfo{person}{Yong~Ho Hwang}.}
  \bibinfo{year}{2021}\natexlab{}.
\newblock \showarticletitle{\#{{Twiti}}: {{Social Listening}} for {{Threat
  Intelligence}}}. In \bibinfo{booktitle}{\emph{Proceedings of the {{Web
  Conference}} 2021}}. \bibinfo{publisher}{Association for Computing
  Machinery}, \bibinfo{pages}{92--104}.
\newblock
\urldef\tempurl%
\url{https://doi.org/10.1145/3442381.3449797}
\showDOI{\tempurl}


\bibitem[Tawil and Alqaraleh(2021)]%
        {tawil_bert_2021}
\bibfield{author}{\bibinfo{person}{Yahya Tawil} {and} \bibinfo{person}{Saed
  Alqaraleh}.} \bibinfo{year}{2021}\natexlab{}.
\newblock \showarticletitle{{{BERT Based Topic-Specific Crawler}}}. In
  \bibinfo{booktitle}{\emph{2021 {{Innovations}} in {{Intelligent Systems}} and
  {{Applications Conference}} ({{ASYU}})}}. \bibinfo{pages}{1--5}.
\newblock
\urldef\tempurl%
\url{https://doi.org/10.1109/ASYU52992.2021.9599076}
\showDOI{\tempurl}


\bibitem[Tundis et~al\mbox{.}(2020)]%
        {tundis_automated_2020}
\bibfield{author}{\bibinfo{person}{Andrea Tundis}, \bibinfo{person}{Samuel
  Ruppert}, {and} \bibinfo{person}{Max M{\"u}hlh{\"a}user}.}
  \bibinfo{year}{2020}\natexlab{}.
\newblock \showarticletitle{On the {{Automated Assessment}} of~{{Open-Source
  Cyber Threat Intelligence~Sources}}}. In
  \bibinfo{booktitle}{\emph{Computational {{Science}} -- {{ICCS}} 2020}},
  \bibfield{editor}{\bibinfo{person}{Valeria~V. Krzhizhanovskaya},
  \bibinfo{person}{G{\'a}bor Z{\'a}vodszky}, \bibinfo{person}{Michael~H. Lees},
  \bibinfo{person}{Jack~J. Dongarra}, \bibinfo{person}{Peter M.~A. Sloot},
  \bibinfo{person}{S{\'e}rgio Brissos}, {and} \bibinfo{person}{Jo{\~a}o
  Teixeira}} (Eds.). \bibinfo{publisher}{Springer International Publishing},
  \bibinfo{pages}{453--467}.
\newblock
\urldef\tempurl%
\url{https://doi.org/10.1007/978-3-030-50417-5_34}
\showDOI{\tempurl}


\bibitem[Tundis et~al\mbox{.}(2022)]%
        {tundis_featuredriven_2022}
\bibfield{author}{\bibinfo{person}{Andrea Tundis}, \bibinfo{person}{Samuel
  Ruppert}, {and} \bibinfo{person}{Max M{\"u}hlh{\"a}user}.}
  \bibinfo{year}{2022}\natexlab{}.
\newblock \showarticletitle{A {{Feature-driven Method}} for {{Automating}} the
  {{Assessment}} of {{OSINT Cyber Threat Sources}}}.
\newblock \bibinfo{journal}{\emph{Computers \& Security}}
  \bibinfo{volume}{113} (\bibinfo{year}{2022}), \bibinfo{pages}{102576}.
\newblock
\urldef\tempurl%
\url{https://doi.org/10.1016/j.cose.2021.102576}
\showDOI{\tempurl}


\bibitem[Vieira et~al\mbox{.}(2016)]%
        {vieira_finding_2016}
\bibfield{author}{\bibinfo{person}{Karane Vieira}, \bibinfo{person}{Luciano
  Barbosa}, \bibinfo{person}{Altigran~Soares Silva}, \bibinfo{person}{Juliana
  Freire}, {and} \bibinfo{person}{Edleno Moura}.}
  \bibinfo{year}{2016}\natexlab{}.
\newblock \showarticletitle{Finding seeds to bootstrap focused crawlers}.
\newblock \bibinfo{journal}{\emph{World Wide Web}}  \bibinfo{volume}{19}
  (\bibinfo{year}{2016}), \bibinfo{pages}{449--474}.
\newblock
\urldef\tempurl%
\url{https://doi.org/10.1007/s11280-015-0331-7}
\showDOI{\tempurl}


\bibitem[Wang et~al\mbox{.}(2010)]%
        {wang_focused_2010}
\bibfield{author}{\bibinfo{person}{Wenxian Wang}, \bibinfo{person}{Xingshu
  Chen}, \bibinfo{person}{Yongbin Zou}, \bibinfo{person}{Haizhou Wang}, {and}
  \bibinfo{person}{Zongkun Dai}.} \bibinfo{year}{2010}\natexlab{}.
\newblock \showarticletitle{A {{Focused Crawler Based}} on {{Naive Bayes
  Classifier}}}. In \bibinfo{booktitle}{\emph{2010 {{Third International
  Symposium}} on {{Intelligent Information Technology}} and {{Security
  Informatics}}}}. \bibinfo{pages}{517--521}.
\newblock
\urldef\tempurl%
\url{https://doi.org/10.1109/IITSI.2010.30}
\showDOI{\tempurl}


\bibitem[Zhang et~al\mbox{.}(2021)]%
        {zhang_dsdd_2021}
\bibfield{author}{\bibinfo{person}{Haoxiang Zhang}, \bibinfo{person}{A{\'e}cio
  Santos}, {and} \bibinfo{person}{Juliana Freire}.}
  \bibinfo{year}{2021}\natexlab{}.
\newblock \showarticletitle{{{DSDD}}: {{Domain-Specific Dataset Discovery}} on
  the {{Web}}}. In \bibinfo{booktitle}{\emph{Proceedings of the 30th {{ACM
  International Conference}} on {{Information}} \& {{Knowledge Management}}}}.
  \bibinfo{publisher}{Association for Computing Machinery},
  \bibinfo{pages}{2527--2536}.
\newblock
\urldef\tempurl%
\url{https://doi.org/10.1145/3459637.3482427}
\showDOI{\tempurl}


\end{thebibliography}

\end{document}